\documentclass[prl,reprint,superscriptaddress]{revtex4-2}

\draft 

\usepackage{siunitx}
\usepackage{amsmath}
\usepackage{amssymb}
\usepackage{braket}
\usepackage{printlen}
\usepackage{graphicx}
\usepackage{float}
\uselengthunit{mm}
\usepackage{hyperref}

\DeclareSIUnit\counts{cps}

\makeatletter
\newcommand{\fontinfo}{%
	Font: \texttt{\fontname\font}, size: \f@size\,pt%
}
\makeatother

\begin{document}
	
	
	\title{Overcoming the Hong-Ou-Mandel interference limitation for the second photon from a two-photon cascade} 
	
	
	
	\author{Raphael Joos}
	\email[]{r.joos@ihfg.uni-stuttgart.de}
	\author{Michal Vyvle\v{c}ka}
	\author{Tim Strobel}
	\author{Benjamin Breiholz}
	\author{Furkan Aglarci}
	\author{Ponraj Vijayan}
	\affiliation{Institut für Halbleiteroptik und Funktionelle Grenzflächen (IHFG), Center for Integrated Quantum Science and Technology (IQ$^{ST}$) and SCoPE, University of Stuttgart, Allmandring 3, 70569 Stuttgart, Germany}
	\author{Hans-Georg Babin}
	\author{Arne Ludwig}
	\affiliation{Faculty of Physics and Astronomy, Experimental Physics VI, Ruhr University Bochum,
		Universitätsstrasse 150, 44801 Bochum, Germany}
	\author{Michael Jetter}
	\author{Simone L. Portalupi}
	\author{Peter Michler}
	\affiliation{Institut für Halbleiteroptik und Funktionelle Grenzflächen (IHFG), Center for Integrated Quantum Science and Technology (IQ$^{ST}$) and SCoPE, University of Stuttgart, Allmandring 3, 70569 Stuttgart, Germany}


	
	
\begin{abstract}
	
Quantum emitters such as semiconductor quantum dots can provide entangled photon pairs emitting from a system of cascaded states. However, the resulting temporal correlation between the two photons from the three-level ladder system has been shown to set a limit on the achievable two-photon interference. In this work, we show that such limitation, for photons emitted by the second transition, is only due to reduced temporal overlap at the interference beam splitter. We investigate this theoretically and experimentally by determining time-resolved four-photon coincidences between two successively emitted photon pairs. Post-emission synchronization shows the recovery of maximum interference visibility; in prospective quantum networks this could be achieved via deterministic quantum memories.



\end{abstract}
	
	

	\maketitle 

\begin{figure}[h]
	\centering
	\includegraphics{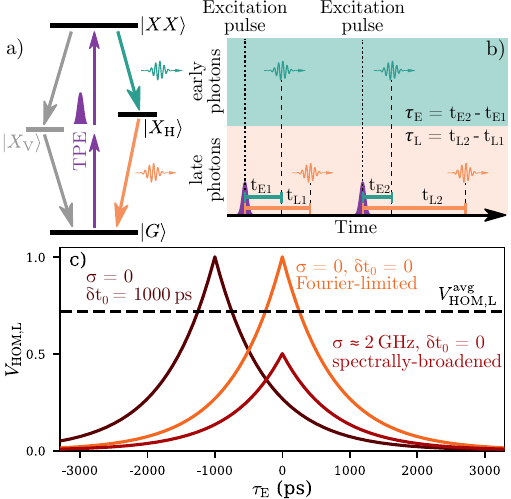}
	\caption{a) Energy level structure of the system under investigation. One decay channel (via $\ket{X_\text{H}}$) of a XX-X-cascade of a QD is selected via polarization filtering resulting in an effective three-level ladder system. The system is prepared in the excited state via fast, resonant two-photon excitation (TPE) with a pulsed laser. b) Exemplarily, two consecutive excitation pulses are shown for the definition of the detection times of the subsequently emitted photons, i.e. the two early and two late photons. c) Calculated HOM visibility of the full wave packet of the late photons dependent on the time delay $\tau_\text{E}$ according to equation \eqref{eq:Vis}.}
	\label{fig:Theory}
\end{figure}

\textit{Introduction}\textemdash Photonic entanglement and indistinguishability are key resources for virtually any operation in advanced quantum networks. For instance quantum repeaters, which will enable large scale networks, can be realized using entanglement swapping between two remote nodes \cite{Azuma2023,VanLoock2020}. Quantum emitters, such as semiconductor quantum dots (QDs), are highly appealing for these applications due to potentially deterministic generation of quantum light \cite{Tomm2021, Ding2025}. Both, near-unity entanglement \cite{Pan2026} as well as indistinguishability \cite{Zhai2022}, have been individually achieved with QD sources. However, the simultaneous generation of such highly entangled and indistinguishable photons has so far been challenging. This is direct consequence of the cascaded emission process via the biexciton-exciton (XX-X) cascade \cite{Simon2005, Scholl2020} which is commonly employed for the generation of polarization-entangled photon pairs with QDs. Due to the finite lifetime of the intermediate state, the early (XX) photon is spectrally broadened leading to reduced Hong-Ou-Mandel (HOM) interference visibility \cite{Chiang2010, Undeutsch2025}. On the other hand, the late (X) photon is temporally correlated with the early photon. Without consideration of these correlations, also the HOM visibility of the late photon is limited to the same value as the early photon \cite{Baltisberger2026}. In this work, we fully resolve these temporal correlations and their impact on the HOM visibility of the late photons. We derive theoretically and measure experimentally a four-photon correlation between two consecutively emitted photon pairs after HOM interference of the late photons. This allows to determine the HOM visibility of late photons, which turns out to be dependent on the actual emission time of the early photons. Ultimately, we show that the reduction of HOM visibility of the late photons is a result of reduced temporal overlap at the interference beam splitter (BS) and that this can be recovered after the emission process reclaiming maximum indistinguishability.\\
\textit{Theory}\textemdash We investigate HOM interference of photons emitted by consecutive emission events of a three-level ladder system, as depicted in figure \ref{fig:Theory} a). The system consists of an upper level $\ket{XX}$, which is coherently excited from the ground state $\ket{G}$ via a two-photon excitation process \cite{Muller2014}, and an intermediate state $\ket{X_\text{H}}$. Experimentally, this system is obtained by polarization-based selection of one decay channel of the XX-X cascade of a semiconductor QD. The transitions $\ket{XX}\rightarrow\ket{X_\text{H}}$ and $\ket{X_\text{H}}\rightarrow\ket{G}$ lead to emission of an early and late photon, respectively, which are described by the two-photon wave packet
\begin{align}
		\xi(t_\text{E}, t_\text{L}) = &\frac{1}{\sqrt{T_\text{E}T_\text{L}}}\Theta(t_\text{E})e^{-t_\text{E}/2T_\text{E}}e^{-i\omega_\text{E}t_\text{E}}\Theta(t_\text{L} -t_{0}-t_\text{E})\nonumber\\
		&e^{-\left(t_\text{L} - t_{0}-t_\text{E}\right)/2T_\text{L}}e^{-i\omega_\text{L}(t_\text{L} -t_{0}-t_\text{E})}.
\end{align}
$T_i$ and $\omega_i$ describe the lifetime and frequency of the photons whereas $t_\text{E}$ ($t_\text{L}$) is the detection time of the early (late) photon after the state $\ket{XX}$ was prepared. $\Theta(t)$ is the heaviside function. In order to perform HOM interference, the system is excited with consecutive pulses (see figure \ref{fig:Theory} b)).
Consequently, $\tau_\text{E}$ ($\tau_\text{L}$) is the difference of the time between excitation pulses and detection of the early (late) photons of two subsequent pulses. Furthermore, we consider an additional delay $t_{0,i}$ of the interfering, late photons after the emission: this results in an temporal offset $\delta t_{0} = t_{0,2} - t_{0,1}$ at the interference BS, which later serves as an additional tool to probe the impact of the temporal correlations. With the help of photon correlation functions \cite{Legero2006, Kambs2018} $\mathcal{G}^{(4)}_{\parallel/\perp}(t_\text{E}, t_\text{E} + \tau_\text{E}, t_\text{L}, t_\text{L} + \tau_\text{L})$ for photons entering the interferometer with identical ($\parallel$) or orthogonal ($\perp$) polarization, we can calculate the total coincidence probability of all four photons after the two late photons interfered
\begin{align}\label{eq:P4}
	\mathcal{P}^{(4)}_{\parallel/\perp} &= \int_{\mathbb{R}^4}\,d\tau_\text{E}\,d\tau_\text{L}\,dt_\text{E}\,dt_\text{L}\, \mathcal{G}^{(4)}_{\parallel/\perp}(t_\text{E}, t_\text{E} + \tau_\text{E}, t_\text{L}, t_\text{L} + \tau_\text{L})\nonumber\\
	& = \int_{\mathbb{R}^2}\,d\tau_\text{E}\,d\tau_\text{L}\,p^{(4)}_{\parallel/\perp}(\tau_\text{E}, \tau_\text{L}).
\end{align}
Physically, only the relative time delays $\tau_\text{E}$ and $\tau_\text{L}$ determine the interference behavior. This leads, via integration over the absolute emission times $t_\text{E}$ and $t_\text{L}$, to the four-photon probability density function
\begin{equation}\label{eq:p4}
	\begin{split}
		p^{(4)}_{\parallel/\perp}(\tau_\text{E}, \tau_\text{L}) =&\frac{1}{16T_\text{E}T_\text{L}}e^{-|\tau_\text{E}|/T_\text{E}} \left(e^{-|\tau_\text{L}+\tau_\text{E}+\delta t_0|/T_\text{L}}\right.\\
		&+e^{-|\tau_\text{L}-\tau_\text{E}-\delta t_0|/T_\text{L}}-2V_{\parallel/\perp}e^{-\sigma^2\tau_\text{L}^2/2}\\
		&\left.e^{-\left(|\tau_\text{L}|+|\tau_\text{E}+\delta t_0|\right)/T_\text{L}}\right).
	\end{split}
\end{equation}
Here, the capability of the late photons to interfere is considered via the interferometric visibility ($V_\parallel = 1$, $V_\perp = 0$), and the distribution in emission frequency differences: the latter is typically determined by the inhomogeneous linewidth $\sigma_\text{inh}$ of the emitter with $\sigma^2 = 2\sigma^2_\text{inh}$ \cite{Kambs2018}. The probability density function fully describes the temporal aspects of the HOM interference considering the temporal correlation between early and late photons. The coincidence probability density of a standard HOM measurement of the late photons is obtained via $p^{(2)}_{\parallel/\perp}(\tau_\text{L}) = \int_{-\infty}^{\infty}\,\text{d}\tau_\text{E}\, p^{(4)}_{\parallel/\perp}(\tau_\text{E}, \tau_\text{L})$. Furthermore, from the coincidence probabilities for indistinguishable and distinguishable photons defined in equation \eqref{eq:P4} the mean HOM visibility can be calculated as
\begin{equation}
	\begin{split}
		V^\text{avg}_\text{HOM,L} = 1 - \frac{\mathcal{P}^{(4)}_\parallel}{\mathcal{P}^{(4)}_\perp} = \frac{1}{2T_\text{E}}\int\limits_{-\infty}^\infty d\tau_\text{E}\,e^{-|\tau_\text{E}|/T_\text{E}}V_\text{HOM,L}(\tau_\text{E}, \delta t_0)
	\end{split}
\end{equation}
with
\begin{equation}\label{eq:Vis}
	\begin{split}
		V_\text{HOM,L}(\tau_\text{E}, \delta t_0) = V_\text{spec}(\sigma, T_\text{L})e^{-|\tau_\text{E} + \delta t_0|/T_\text{L}}.
	\end{split}
\end{equation}
Here, $V_\text{HOM,L}$ is the HOM visibility of the full wave packet of independent photons (no temporal correlations due to cascaded emission) with temporal offset $\tau_\text{E} + \delta t_0$ at the input of the interferometer. A reduced spectral overlap of the photons is considered via $V_\text{spec}(\sigma, T_\text{L})$\cite{Kambs2018} which takes a value of one for Fourier-limited photons. $V_\text{spec}(\sigma, T_\text{L})$ leads to an overall reduction of the HOM visibility independent of the temporal aspects of the interference. A more detailed description of the derivation of the probability density function and the time-resolved HOM visibility can be found in the supplemental material (SM). Interestingly, the visibility of a standard HOM measurement, here $V^\text{avg}_\text{HOM,L}$, is nothing more than the average of $V_\text{HOM,L}(\tau_\text{E}, \delta t_0)$ weighted by the distribution of input delays given by the exponential decay of the early photons. For $\delta t_0= 0$ and $\sigma = 0$, the well-known result of $V^\text{avg}_\text{HOM,L} = 1/(1+T_\text{E}/T_\text{L})$ \cite{Scholl2020} is obtained. Figure \ref{fig:Theory} c) shows the resulting time-resolved visibility $V_\text{HOM,L}$ of the late photons for three different scenarios of $\delta t_0$ and $\sigma$. All simulations use values of $T_\text{E} = \SI{294}{\pico\second}$ and $T_\text{L} = \SI{757}{\pico\second}$ which are the same as for QD1 used in the following experiment (QD XX and X lifetime characterization can be found in the SM). Fourier-limited ($\sigma = 0$) and synchronized ($\delta t_0 = 0$) photons reach unity visibility for $\tau_\text{E} = 0$. However, for increasing $\tau_\text{E}$ the visibility decreases rapidly (below the average value) ultimately converging towards zero for $\tau_\text{E} \gg T_\text{L}$. In case of decreased photon coherence ($\sigma>0$), the visibility curve is vertically compressed. The maximum visibility value is solely determined by the degree of spectral broadening of the emitter. Interestingly, an additional input delay before the interferometer (here $\delta t_0 = \SI{1000}{\pico\second}$) does not simply lead to a reduction of the visibility. In particular, unity visibility ($\sigma = 0$) is obtained for $\tau_\text{E} = \SI{-1000}{\pico\second}$ and the entire curve is horizontally shifted by said value. In this case, the additional input delay compensates the initial temporal offset between two late photons (when $\tau_\text{E} = \SI{-1000}{\pico\second}$) leading to perfect temporal overlap and recovering maximum indistinguishability. This implies that HOM interference of the late photons is not intrinsically limited due to the cascaded emission process, as it is the case for the spectrally broadened early photons \cite{Chiang2010,Undeutsch2025}. Essentially, interference of the late photons reduces to a matter of synchronization at the interferometer.\\
\textit{Results}\textemdash
\begin{figure}[b]
	\centering
	\includegraphics{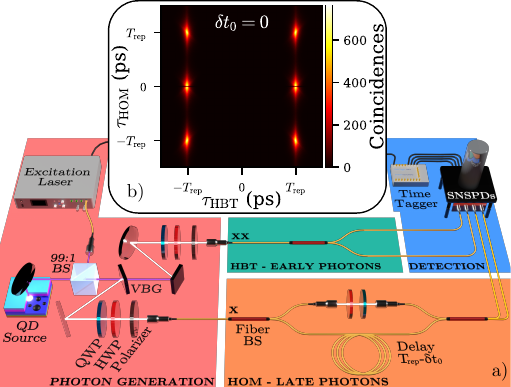}
	\caption{a) Sketch of the experimental setup for the measurement of time-resolved HOM. The XX-X-cascade of a QD is excited via resonant TPE with a repetition period of $T_\text{rep} \approx \SI{6.6}{\nano\second}$. The early (XX) and late (X) photons are separated via volume Bragg grating (VBG) filters and projected onto H-polarizers. Subsequently, the early photons enter a Hanbury-Brown and Twiss (HBT) interferometer while late photons of consecutive pulses undergo HOM inteference in an unbalanced Mach-Zehnder interferometer. QWP and HWP in one interferometer arm enable parallel and orthogonal polarization setting at the interference BS. The detection events of superconducting nanowire single-photon detectors (SNSPDs) and the reference clock of the excitation laser meet at a TimeTagger. b) Two-dimensional histogram of the four-photon coincidence events for parallel polarization. The correlation is conditioned on a simultaneous detection of one early-late photon pair within the same repetition period. This is crucial and ensures that the coordinate of (0,0) corresponds to simultaneous detection of all four photons.}
	\label{fig:Setup}
\end{figure}
In order to verify the theoretical findings, we perform measurements of the probability density function $p^{(4)}_{\parallel/\perp}$ which allows to fully characterize the time dependence of $V_\text{HOM,L}$ of the late photons. The experimental details follow exactly the outlined structure of the schematic in figure \ref{fig:Theory} a), b) and figure \ref{fig:Setup} a) outlines the technical realization of the experiment. In a first step, we use a high-efficiency source of entangled photon pairs (QD1) in order to thoroughly investigate the temporal profile of $p^{(4)}_{\parallel/\perp}$. QD1 is an InAs QD on a metamorphic buffer layer incorporated in a planar cavity structure emitting in the telecom C-band. An in-depth characterization of the source can be found in reference \cite{Joos2026}. We condition all coincidence events on the simultaneous detection of one early-late (XX-X) photon pair within the same repetition period. By doing so, an intuitive two-dimensional histogram (see figure \ref{fig:Setup} b) for $\delta t_0 = 0$) can be obtained (instead of a three-dimensional histogram as typically given for any $G^{(4)}$ measurement) where the time coordinate (0,0) corresponds to simultaneous detection of all four photons. Due to the single-photon nature of the early photons, the peaks of the center column (around $\tau_\text{HBT}= 0$) are vanishing while the HOM measurement of the late photons interference gives rise to a dip around the center row ($\tau_\text{HOM}= 0$). For the probability density function $p^{(4)}_{\parallel/\perp}$, the two time delays $\tau_\text{E}$ and $\tau_\text{L}$ are required. The latter is directly given by the time delay of the HOM measurement (y-axis). For the HBT measurement, two consecutive photons are naturally at the first positive (negative) sidepeak (x-axis)  with $\tau_\text{E} = \tau_\text{HBT} - T_\text{rep}$ ($\tau_\text{E} = -\tau_\text{HBT} - T_\text{rep}$). The measurement of the probability density function of parallel and orthogonal photons is performed for two input delays of $\delta t_0 = 0$ and $\delta t_0 = \SI{300}{\pico\second}$.
\begin{figure}[t]
	\centering
	\includegraphics{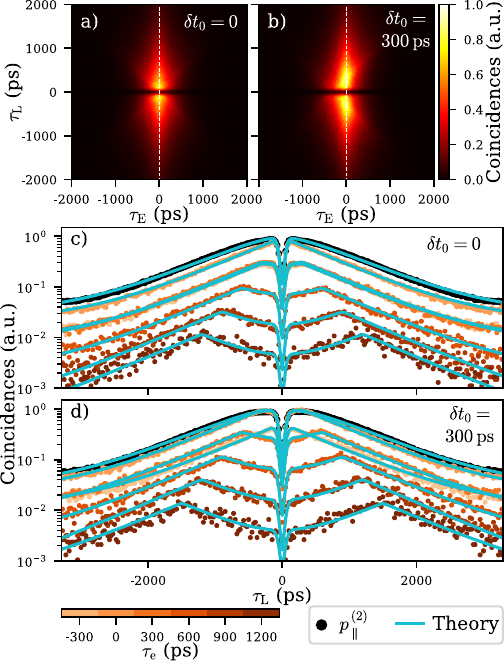}
	\caption{Measured probability density function $p^{(4)}_{\parallel}(\tau_\text{E}, \tau_\text{L})$ for QD1 with a) $\delta t_0 = 0$, b)  $\delta t_0 = \SI{300}{\pico\second}$. c), d) Corresponding cross-sections for different values of $\tau_\text{E}$ as well as the standard HOM measurement $p^{(2)}_{\parallel}(\tau_\text{L})$ (black curve). Cross-sections at $\tau_\text{E} = 0$ correspond to the white dotted lines in a), b).  Normalization is performed so that $p^{(2)}_{\perp}(0) = p^{(4)}_{\perp}(0, \delta t_0) = 1$.}
	\label{fig:Correlations}
\end{figure}
For both delays, figure \ref{fig:Correlations} a) and b) show the full two-dimensional histogram for parallel polarization while c) and d) depict the corresponding cross-sections (vertical cuts) for different values of $\tau_\text{E}$, respectively. A fit to $p^{(2)}_{\parallel}$ yields the broadening parameter $\sigma$ which is then used to theoretically model the behavior of  $p^{(4)}_{\parallel}$. Fit and simulation include a convolution with the binwidth of $\tau_\text{E}$ of \SI{75}{\pico\second} and two-channel detector resolution of \SI{50}{\pico\second}. The values for the measurement with $\delta t_0 = 0$ ($\sigma = \SI{21.8\pm0.2}{\giga\hertz}$) and $\delta t_0 = \SI{300}{\pico\second}$ ($\sigma = \SI{18.8\pm0.2}{\giga\hertz}$) differ slightly which can be caused by a change in the charge environment of QD1 during the measurement. Overall, the data show an excellent agreement with the theoretically predicted behavior. The interference dip is always centered around $\tau_\text{L} = 0$ as expected for any HOM measurement. Without additional input delay (figure \ref{fig:Correlations} a), c)), the overlap of the two late photons decreases continuously with increasing $|\tau_\text{E}|$ which can be seen as follows. The cross-section for $\tau_\text{E} = 0$ is identical to the result of a HOM measurement with single, uncorrelated photons without cascaded emission. However, for larger values of $|\tau_\text{E}|$, two increasingly separate peaks at $\pm\tau_\text{E}$ are present (cf. equation \eqref{eq:p4}) signifying the reduction of the overlap between the two late photons. However, with $\delta t_0 = \SI{300}{\pico\second}$ (figure \ref{fig:Correlations} b), d)), the signal is no longer symmetric around $\tau_\text{E} = 0$. Now, the best overlap (cross-section as for single, uncorrelated photons) is obtained for late photons which originate from emission events of $\tau_\text{E} = \SI{-300}{\pico\second}$. Consequently, synchronization of these photons is achieved after emission with initial temporal mismatch. 
Finally, we determine the HOM visibility of the late photons from the measurements of $p^{(4)}_{\parallel}$. For each value of $\tau_\text{E}$, we integrate all counts of the cross-section of $p^{(4)}_{\parallel}$ within one repetiton period ($\pm\,\SI{3292}{\pico\second}$, full window in figure \ref{fig:Correlations} c), d)) yielding $N_\parallel$. As reference for distinguishable photons, we use the theoretical model of $p^{(4)}_{\perp}$ based on the fits depicted in figure \ref{fig:Correlations} which gives $N_\perp$. Any HOM visibility is then calculated as $V = 1-\frac{N_\parallel}{N_\perp}$. The same analysis is performed for $p^{(2)}_{\parallel}$ resulting in the corresponding value of $V^\text{avg}_\text{HOM,L}$. A direct comparison of the measurements with parallel and orthogonal photons can be found in the SM. The latter method yields the same qualitative behavior but is more prone to errors in the existing experimental configuration. 
\begin{figure*}[t]
	\centering
	\includegraphics{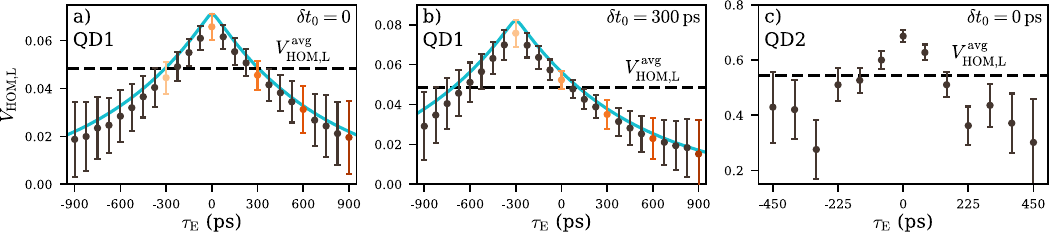}
	\caption{Time-resolved HOM visibility of the late photons a) $V_\text{HOM,L}(\tau_\text{E}, 0)$  and  b) $V_\text{HOM,L}(\tau_\text{E}, \SI{300}{\pico\second})$ for QD1. Theory curve depicted by the cyan line. The colored dots correspond to the visibility for $\tau_\text{E}$ of the equally colored curves in figure \ref{fig:Correlations}. c) $V_\text{HOM,L}(\tau_\text{E}, 0)$ for QD2, a droplet-etched GaAs QD in a p-i-n diode structure.}
	\label{fig:Visibility}
\end{figure*}
Figure \ref{fig:Visibility} a) and b) show the results for the two input delays of $\delta t_0 = 0$ and $\delta t_0 = \SI{300}{\pico\second}$, respectively. Both measurements show the anticipated behavior from equation \eqref{eq:Vis} with an exponentially decreasing visibility for decreasing photon overlap whereas maximum visibility is found for the condition of $\tau_\text{E} = -\delta t_0$. The latter values of $V_\text{HOM,L}(0,0) = \SI{6.6\pm0.5}{\percent}$ and $V_\text{HOM,L}(\SI{300}{\pico\second},\SI{-300}{\pico\second}) = \SI{7.6\pm0.7}{\percent}$ clearly exceed the standard HOM visibilities of $V^\text{avg}_\text{HOM,L}$ of \SI{4.8\pm0.3}{\percent}/\SI{4.9\pm0.3}{\percent}, respectively. The maximum values of $V_\text{HOM,L}$ should be the same for both input delays. The slightly higher value for $\delta t_0 = \SI{300}{\pico\second}$ is, in turn, due to the reduced spectral broadening (lower $\sigma$ value) during that measurement. While QD1 shows an excellent agreement with the theoretically predicted temporal behavior, the overall HOM visibility is significantly limited by spectral broadening of the emission line. In order to unanimously verify the findings of this work, we perform the same experiment (however, with  $T_\text{rep} \approx \SI{2.6}{\nano\second}$) with a second emitter, as depicted in figure \ref{fig:Visibility} c). QD2 is a droplet-etched GaAs QD emitting at about \SI{780}{\nano\meter} in a p-i-n diode structure as described in reference \cite{Chen2024}. In good agreement to the results of QD1, the visibility is increasing for lower values of $|\tau_\text{E}|$ culminating in $V_\text{HOM,L}(0,0) = \SI{69\pm2}{\percent}$. The latter value clearly exceeds the measured standard HOM visibility of $V^\text{avg}_\text{HOM,L} = \SI{54.4\pm1.5}{\percent}$ and even surpasses the threshold for Fourier-limited photons of $V^\text{avg}_\text{HOM,L} = 1/(1+\SI{152}{\pico\second}/\SI{275}{\pico\second}) = \SI{64.4}{\percent}$ given by the lifetimes of the states $\ket{XX}$ and $\ket{X_\text{H}}$ (see SM). Due to the lower signal-to-noise ratio of QD2, no reliable fit to the data could be performed. Hence, $N_\perp$ is taken from the reference measurement with orthogonal polarization (see SM). The remaining limitation of $V_\text{HOM,L}(0,0)$ is most likely due to the non-zero spectral broadening of the emission line. This is backed up by an additional HOM measurement of the late photons of QD2 under longitudinal-acoustic phonon-assisted (LA) excitation of the $\ket{X_\text{H}}$ state. In the absence of spectral broadening, this excitation scheme promises close to unity HOM visibility \cite{Gustin2020} which is experimentally comparable to the values obtained under resonant excitation \cite{Reindl2019}. Here, we achieve a HOM visibility of \SI{73.4\pm0.3}{\percent} under LA excitation (see SM) which is close to $V_\text{HOM,L}(0,0)$ indicating that also the latter is only limited by broadening of the line and no further effect due to the cascaded emission. The findings of this work also lead to implications for similar effects for other excitation schemes. Without measurement of the early photon, the cascaded emission process is the very same as incoherent excitation of the intermediate state and subsequent emission. Therefore, also other excitation mechanisms of quantum emitters, which limit the interference due to temporal jitter, should have the same fingerprint as the time-resolved $V_\text{HOM,L}$ presented here. This comprises for instance excitation via relaxation processes from higher electronical \cite{Reindl2019, Hauser2026} or vibrational \cite{Schofield2022, Trebbia2010} states. A direct measurement of the time-resolved $V_\text{HOM,L}$ would, however, be arguably harder to measure due to the absence of a photon heralding the exact time of the relaxation process.\\
\textit{Conclusion}\textemdash We have theoretically derived and experimentally demonstrated the origin of the limitation of HOM interference of the late photon from a three-level ladder system: the variance in the emission time of the early photon leads, in average, to a temporal mismatch of the two late photons at the interference BS. Photons with perfect temporal overlap exhibit, however, no degradation in interference visibility other than natural decoherence mechanisms of the intermediate state itself. The temporal overlap can also be obtained by post-emission synchronization of late photons which initially feature a temporal offset due the cascaded emission. This becomes especially interesting in the context of quantum networks. Typically, quantum memories are required for the synchronization of different nodes of the network. Consequently, these quantum memories \cite{Gundogan2012, Makino2016,Thomas2024,Gera2024} can be also used as a natural tool to eliminate the timing jitter due to the cascaded emission (or other potential excitation mechanisms including relaxation) and, thus, restore maximum interference visibility. This comes (for the late photons) practically without additional costs, as for instance complex, imbalanced cavity structures \cite{Baltisberger2026, Behrends2026}. Together, this fundamental study denotes an important step towards the practical implementation of quantum networks with photons from cascaded emission. Especially for QDs, this enables simultaneous exploitation of entanglement and indistinguishability from photons of the XX-X-cascade. This facilitates for instance efficient entanglement swapping processes between two entangled photon pairs, a key requisite of many quantum repeater architectures \cite{Azuma2023,VanLoock2020}.\\\\
\textit{Acknowledgments}\textemdash R.J., M.V., T.S., B.B., F.A., P.V., M.J., S.L.P. and P.M. acknowledge funding  by the German Federal Ministry of Research, Technology and Space (BMFTR) via project QR.X (16KISQ013) and QR.N (16KIS2207). Additional funding was also provided via the project EQSOTIC. This project was funded within the QuantERA II Programme that has received funding from the EU’s H2020 research and innovation programme under the GA No 101017733, and with funding organisation BMFTR (with project number 16KIS2060K). H.G.B. and A.L. acknowledge funding by the BMFTR via the projects EQSOTIC No. 16KIS2061, and QR.N No. 16KIS2200.

\bibliography{References}
	
\end{document}